\documentclass[letterpaper, 10 pt, conference]{ieeeconf}  

\IEEEoverridecommandlockouts                              
\usepackage{amsmath}
\usepackage{amsfonts}
\usepackage{amssymb}
\newtheorem{lem}{Lemma}
\newtheorem{stab}{Theorem}
\newtheorem{rem}{Remark}

\title{\LARGE \bf  Stability Analysis of Discrete-time Lur'e Systems with Slope-restricted Odd Monotonic Nonlinearities
}

\author{Kwang Ki Kevin Kim and Richard D. Braatz
\thanks{This work was supported by the National Science Foundation under Grant \#0426328}
\thanks{K. Kim is with the Department of Aerospace Engineering,
	University of Illinois at Urbana-Champaign, Urbana, IL, 61801, USA
	{\tt\small kkim78@illinois.edu}}%
\thanks{R. Braatz is with the University of Illinois at Urbana-Champaign, Urbana,
	IL, 61801, USA {\tt\small braatz@illinois.edu.}}%
}

\begin{document}

\maketitle

\begin{abstract}
Many nonlinear dynamical systems can be written as Lur'e systems, which are described by a linear time-invariant system interconnected with a diagonal static sector-bounded nonlinearity. Sufficient
conditions are derived for the global asymptotic stability analysis of discrete-time Lur'e systems in which the
nonlinearities have restricted slope and/or are odd, which is the usual case in real applications.
A Lur'e-Postnikov-type Lyapunov function is proposed that is used to derive sufficient analysis conditions
in terms of linear matrix inequalities (LMIs).
The derived stability critera are provably less conservative than criteria published in the literature,
with numerical examples indicating that conservatism can be reduced by orders of magnitude.
\end{abstract}

\section{Introduction}

Absolute stability theory considers the stability of a nominal linear time-invariant system
interconnected with a static nonlinearity, that is, a real function of real variables with
no internal states. The literature on absolute stability theory can be categorized in terms
of the properties of the static nonlinearities that are considered.
This paper derives new sufficient conditions for the stability analysis of discrete-time systems
with the advance being LMI conditions that exploit the static, sector-bounded, and
slope-restricted nature of the componentwise nonlinear operators.

Lyapunov methods provide a simple but powerful way to analyze nonlinear dynamical systems and design
stabilizing controllers~\cite{BGFB94,CheW95,KapH96,KonK99,YZZ07}.
Many well-known results on absolute stability were developed for a benchmark problem known as {\em the Lur'e problem}~\cite{Mey65,NarT73,Par02,ShaS81}.
The Popov and Circle criteria are sufficient frequency-domain conditions for absolute stability
for the feedback interconnection of a continuous linear time-invariant system with a sector-bounded nonlinearity~\cite{GupJ94,HadB94a,Kha02,LeeP08,Sin84,ZamF68}.
Its discrete-time counterparts are known as the Tsypkin criterion~\cite{LarK01,ParK98}
and the Jury-Lee criterion~\cite{JurL64}.
Consider the nonlinear discrete-time dynamical system:
\begin{equation}
\label{eq:1}
\left.
\begin{array}{rl}
x(k+1) = Ax(k) + B p(k),& \\
q(k) = Cx(k) + D p(k),&  p(k) = -\phi(q(k)),
\end{array}
\right.
\end{equation}
where $A \in \mathbb{R}^{n\times n}$, $B \in \mathbb{R}^{n \times n_p}$, $C \in \mathbb{R}^{n_q \times n}$,
$D \in \mathbb{R}^{n_q \times n_p}$, and the nonlinear operator $\phi \in \Phi$, where
$\Phi$ is a set of static functions that satisfy $\phi(0) \equiv 0$ and have some specified input-output characteristics.
This paper considers several classes of diagonal nonlinear operators $\Phi$ that are defined in Appendix \ref{appd:nonlinearftn}, in which the individual nonlinearities are sector-bounded, slope-restricted, and/or odd.

Suppose that the state-space system representation $(A,B,C,D)$ is a minimal realization of the transfer function $G(z)=C(zI - A)^{-1}B +D$
such that the triplet $(A,B,C)$ is controllable and observable.\footnote{Alternatively, the controllability and observability conditions can be removed because strict inequalities will be used
in the stability criteria (see Section 2 in~\cite{Ran96}). Instead, the condition $\mbox{Ker}A \cap \mbox{Ker}B = \{0\}$, which implies that the origin is the one equilibria, is assumed to hold.}
It is well known that the frequency-domain sufficient stability conditions---the Circle, Popov and Tsypkin criteria---for the absolute stability of the system (\ref{eq:1}) can be simply represented by linear matrix inequalities (LMIs) written in terms of the matrices $\{A,B,C,D\}$ via the Positive Real Lemma and the KYP lemma~\cite{BGFB94,ParK98,YZZ07}.  In this paper the S-procedure converts the constraints on the nonlinearities
into LMI conditions, in the same way as in the direct derivation of LMI-based tests for absolute stability~\cite{BGFB94,MegT93,Par02,Yak92}.\footnote{Related results have been derived based on integral quadratic constraints (IQCs)~\cite{DRMJ01,JonM98,Jon97,MegR97,MegT93,Ran96}
for systems with repeated monotonic or slope-restricted nonlinearities in continuous-time cases.}
A modified Lur'e-Postnikov function is proposed that produces less conservative conditions for global asymptotic stability.  The resulting LMI problems are solved for several numerical examples using off-the-shelf software~\cite{GNLC95,SeDuMi,YALMIP}.

\section{Sufficient Conditions for Asymptotic Stability}
\label{sec:suffstabmain}

\subsection{Modified Lur'e-Postnikov Stability of a Discrete-time Lur'e System}

Due to the conservativeness inherent to quadratic Lyapunov functions when applied to nonlinear systems,
this section considers the use of a modified Lur'e-Postnikov function that includes integral terms whose
sign definiteness are implied by the sector-bounded property of the nonlinear operators ($\phi \in \Phi_{sb}^{[0,\xi]}$).
To derive sufficient conditions for the origin of (\ref{eq:1}) to be globally asymptotically stable,
consider the Lyapunov function:
\begin{align}
\label{eq:lyap_general}
 V(x_k) & =
        \bar{x}_k^T
        P
        \bar{x}_k
        + 2 \sum_{i=1}^{n_q} Q_{ii} \int_{0}^{q_{k,i}} \phi_i(\sigma) d\sigma \nonumber \\
        & + 2 \sum_{i=1}^{n_q} \tilde{Q}_{ii} \int_{0}^{q_{k,i}} \left[\xi_i \sigma -\phi_i(\sigma)  \right] d\sigma,
\end{align}
where
\begin{align*}
&
\bar{x}_k \triangleq \begin{bmatrix}x_k \\ p_k\\ q_k \end{bmatrix}, \,\,
P^T = P \triangleq
\begin{bmatrix} P_{11} & P_{12} & P_{13} \\ P_{12}^T & P_{22} & P_{23} \\ P_{13}^T & P_{23}^T & P_{33} \end{bmatrix} \geq 0, \,\,
P_{11} > 0 , \\
&
Q_{ii} \geq 0, \,\,
\tilde{Q}_{ii} \geq 0, \,\, \forall i=1,\ldots,n_q,
\end{align*}
and the subscript $k$ indicates a sampling instance.
Both $p_k$ and $q_k$ are functions of the state variable vector $x_k$,
and the above Lyapunov function is radially unbounded and positive for all nonzero $x_k \in \mathbb{R}^n$.

For $P_{12}=0, P_{13}=0, P_{22}=0, P_{23}=0, P_{33}=0 ,
\tilde{Q} \triangleq \mbox{diag}\{\tilde{Q}_{ii}\} = 0$,
the Lyapunov function (\ref{eq:lyap_general}) reduces to the Lur'e-Postnikov
function used in the derivation of the Popov criterion.
All other Lyapunov functions considered in the Lur'e problem literature are subsets of (\ref{eq:lyap_general}).
Some papers in the Lur'e stability literature that appear to have additional terms
actually just introduce the S-procedure in a different way.
The dependence of the Lyapunov function (\ref{eq:lyap_general}) on the nonlinearities and the states
provide additional degrees of freedom for reducing conservatism during stability analysis.
The application of the S-procedure provides a standard way to consider more restrictive classes of nonlinearities
without having modify the Lyapunov function.
The degree of conservatism will be compared for sub-problems with different
subsets of the Lyapunov function (\ref{eq:lyap_general}), to assess the reduction
in conservatism posed by each term.

\subsection{Lagrange Relaxations}

Yakubovich showed that the positiveness of a quadratic function $f_0(x)$ in a constraint set
expressed in terms of quadratic functions, say $f_i(x)$, $i=1,\ldots,m$, can be implied by the
relaxed form with (Lagrange) multipliers.
The S-procedure is a special case of Lagrange relaxation in which the constraints are represented
in terms of quadratic functions, so that the multipliers can be combined into an LMI inequality.
The application of the S-procedure appears not only in control theory but also in the more generic optimization literature. For convenience, below is the form of the S-procedure used in the proofs of this paper.

\begin{lem}(S-procedure for Quadratic Inequalities) For the Hermitian matrices $\Theta_i$, $i=0,\ldots,m$, consider the two sets:\\
(S1) $\zeta^* \Theta_0 \zeta < 0, \ \ \forall \zeta \in \Phi \triangleq \{\zeta \in \mathbb{F}^n| \zeta^* \Theta_i \zeta \leq 0, \,\, \forall i=1, \ldots, m\}$; \\
(S2) $\exists \tau_i \geq 0$, $i=1,\ldots,m$ such that $\Theta_0 - \sum_{i=1}^m \tau_i \Theta_i < 0$.\\
The feasibility of (S2) implies (S1).
\end{lem}

\subsection{Discrete-time Lur'e Systems with Slope-restricted Nonlinearities}

While conic-sector bounded nonlinearities, which bound the global slope of the nonlinear functions $\phi(\cdot)$, have been heavily
studied in the literature, such a description allows the local slope of the nonlinear functions $\phi(\cdot)$ to vary arbitrarily
from one time instance to another. That is, such nonlinearities do not imposes any {\em local} slope restriction.
Most nonlinearities in practice have a local slope restriction,
in which case a less conservative analysis condition may be achieved if these
constraints on the nonlinearities are included in the analysis via the S-procedure.
A local slope restriction also provides an upper bound on the change of the integral term in the Lyapunov function,
provided that $\phi(\cdot)$ is continuous almost everywhere (a.e.).
\begin{stab}
\label{stab:sufflyap_sr}
Consider a system of the form (\ref{eq:1}) with the memoryless nonlinearity $\phi \in \Phi_{sb}^{[0,\xi]} \cap \Phi_{sr}^{[0,\mu]}$
that is continuous almost everywhere. A sufficient condition for global asymptotic stability is the existence
of a positive semidefinite matrix $P=P^T$ with a positive definite submatrix $P_{11}=P_{11}^T$ and diagonal positive semidefinite matrices
$Q$, $\tilde{Q}$, $T$, $\tilde{T}$, $N \in \mathbb{R}^{n_q \times n_q}$ such that
\begin{equation}
G =
\begin{bmatrix}
G_{11} & G_{12} & G_{13} \\
G_{12}^T & G_{22} & G_{23} \\
G_{13}^T& G_{23}^T & G_{33}
\end{bmatrix}
< 0
\label{eq:G1}
\end{equation}
where
\begin{align*}
G_{11} & =
    A^T (P_{11} + P_{13} C + C^T P_{13}^T + C^T P_{33} C) A
    - P_{11}
    \nonumber \\
    &
    - P_{13} C - C^T P_{13}^T
    - C^T P_{33} C + A^T C^T \tilde{Q} \xi C A
    \nonumber \\
    &
    - C^T \tilde{Q} \xi C,
    \\
G_{12} & =
    A^T (P_{11} + P_{13} C + C^T P_{13}^T + C^T P_{33} C) B
    - P_{12}
    \nonumber \\
    &
    - P_{13} D - C^T P_{23}^T - C^T P_{33} D - C^T T
    \nonumber \\
    &
    + (CA - C)^T Q + A^T C^T \tilde{Q} \xi C B +  (CA - C)^T \tilde{Q}
    \nonumber \\
   &
    - C^T \tilde{Q} \xi D + (CA - C)^T \mu N
    \\
G_{13} & =
    A^T P_{12} + A^T P_{13} D + A^T C^T P_{23}^T + A^T C^T P_{33} D
    \nonumber \\
    &
    - A^T C^T \tilde{T} - (CA -C)^T Q + A^T C^T \tilde{Q} \xi D
    \nonumber \\
    &
    - (CA - C)^T \mu N
    \\
G_{22} & =
    B^T (P_{11} + P_{13} C + C^T P_{13}^T + C^T P_{33} C) B
    - P_{22}
    \nonumber \\
    &
    - P_{23} D - D^T P_{23}^T - D^T P_{33} D
    -2 \xi^{-1} T - TD
    \nonumber \\
    &
    -D^T T + Q(CB -D) + (CB -D)^T Q-\mu^{-1} Q
    \nonumber \\
    &
    + B^T C^T \tilde{Q} \xi C B - D^T \tilde{Q} \xi D
    \nonumber \\
    &
    + (CB -D)^T \tilde{Q} + \tilde{Q} (CB-D) - \mu^{-1} \tilde{Q}
    - 2 N
    \nonumber \\
    &
    + N \mu (CB-D)+ (CB-D)^T \mu N
    \\
G_{23} & =
    B^T P_{12} + B^T P_{13} D + B^T C^T P_{23}^T + B^T C^T P_{33} D
    \nonumber \\
    &
    - B^T C^T \tilde{T} + Q D + Q \mu^{-1} + \mu^{-1} \tilde{Q}
    \nonumber \\
   &
    + \tilde{Q} D + B^T C^T \tilde{Q} \xi D + 2 N
    \nonumber \\
    &
    - (CB-D)^T \mu N + N \mu D
    \\
G_{33} & =
    P_{22} + P_{23} D + D^T P_{23}^T + D^T P_{33} D
    -2 \xi^{-1} \tilde{T} -\tilde{T} D
    \nonumber \\
    &
    -D^T \tilde{T} -Q D - D^T Q -Q \mu^{-1}
    - \mu^{-1} \tilde{Q} + D^T \tilde{Q} \xi D
    \nonumber \\
    &
    - 2N - N \mu D - D^T \mu N
\end{align*}
\end{stab}
\noindent
{\em Proof: } The difference in the Lyapunov function (\ref{eq:lyap_general}) between the $k+1$ and $k$ sampling instances is
\begin{align}
\label{eq:del_lyap1}
\Delta V(x_{k}) & =  \zeta_k^T ( A_a^T P A_a - E_a^T P E_a ) \zeta_{k}
                    \nonumber \\
                & + \ 2 \displaystyle \sum_{i=1}^{n_q} Q_{ii} \int_{q_{k,i}}^{q_{k+1,i}} \phi_i(\sigma) d\sigma \\
                & + \ 2 \displaystyle \sum_{i=1}^{n_q} \tilde{Q}_{ii} \int_{q_{k,i}}^{q_{k+1,i}} \left[\xi_i \sigma -  \phi_i(\sigma) \right] d\sigma , \nonumber
\end{align}
where
$$
\zeta_k \triangleq
\begin{bmatrix}
x_k \\ p_k \\p_{k+1}
\end{bmatrix} \!\!,
A_a \triangleq
\begin{bmatrix}
A & B & 0 \\
0 & 0 & I \\
CA & CB & D
\end{bmatrix} \!\!,
E_a \triangleq
\begin{bmatrix}
I & 0 & 0 \\
0 & I & 0 \\
C & D & 0
\end{bmatrix}
$$
Slope restrictions on the nonlinearities place an upper bound on the first integral:
\begin{align*}
& 2 \sum_{i=1}^{n_q} Q_{ii} \int_{q_{k,i}}^{q_{k+1,i}} \phi(\sigma) d\sigma \\
& \leq
2 \sum_{i=1}^{n_q} Q_{ii}
\left\{
\begin{array}{@{}c@{}}
(\phi_{k+1,i}-\phi_{k,i})(q_{k+1,i}-q_{k,i}) \\
- \frac{1}{2 \mu_i}(\phi_{k+1,i}-\phi_{k,i})^2
\end{array}
\right\} \\
& =
\zeta_k^T U_1 \zeta_k,
\end{align*}
where $U_1$ is given in Appendix \ref{appd:matxSproc}.
Similarly, an upper bound can be derived on the second integral:
\begin{align*}
& 2 \sum_{i=1}^{n_q} \tilde{Q}_{ii} \int_{q_{k,i}}^{q_{k+1,i}} \left[\xi_i \sigma -\phi(\sigma)  \right] d\sigma \\
& =
    - 2 \sum_{i=1}^{n_q} \tilde{Q}_{ii} \int_{q_{k,i}}^{q_{k+1,i}} \phi(\sigma) d\sigma
    + 2 \sum_{i=1}^{n_q} \tilde{Q}_{ii} \int_{q_{k,i}}^{q_{k+1,i}} \xi_i \sigma d\sigma \\
& \leq
    - 2 \sum_{i=1}^{n_q} \tilde{Q}_{ii} \left\{ \frac{1}{2 \mu_i} (\phi_{k+1,i}-\phi_{k,i})^2 + \phi_{k,i}(q_{k+1,i}-q_{k,i}) \right\} \\
& + \sum_{i=1}^{n_q} \tilde{Q}_{ii} \xi_i \left[ q_{k+1,i}^2 - q_{k,i}^2 \right] \\
& =
    \zeta_k^T U_2 \zeta_k,
\end{align*}
where $U_2$ is given in Appendix \ref{appd:matxSproc}.

Since the (negative) feedback-connected nonlinearity is monotonic with slope restriction in addition
to being $[0,\xi]$ sector-bounded, i.e., $\phi \in \Phi_{sb}^{[0,\xi]} \cap \Phi_{sr}^{[0,\mu]} $,
it can be shown that the following inequalities are satisfied at each sampling instance $k$ and all indices $i = 1,\ldots,n_q$:
\begin{align}
 \phi_{k,i}[\xi_i^{-1} \phi_{k,i} - q_{k,i}] & \leq 0,
 \label{eq:sect_ineq}\\
 (\phi_{k+1,i} - \phi_{k,i})[\mu_i^{-1}(\phi_{k+1,i} - \phi_{k,i}) - (q_{k+1,i} - q_{k,i}) ] & \leq 0.
\label{eq:slope_ineq}
\end{align}
The following notations based on (\ref{eq:sect_ineq}) are useful when applying the S-procedure:
\begin{align}
\sum_{i=1}^{n_q} 2 \tau_{i} \phi_{k,i}[\xi_i^{-1} \phi_{k,i} - q_{k,i}]
& = \zeta_k^T S_1 \zeta_k,
\\
\sum_{i=1}^{n_q} 2 \tilde{\tau}_{i} \phi_{k+1,i}[\xi_i^{-1} \phi_{k+1,i} - q_{k+1,i}]
& = \zeta_k^T S_2 \zeta_k,
\end{align}
where $S_1$ and $S_2$ are given in Appendix \ref{appd:matxSproc}.
A similar notation based on the inequality (\ref{eq:slope_ineq}) is:
\begin{align}
& \sum_{i=1}^{n_q} 2 N_{ii}(\phi_{k+1,i} - \phi_{k,i}) \left[ \mu_i^{-1}(\phi_{k+1,i} - \phi_{k,i}) -(q_{k+1,i} - q_{k,i}) \right] \nonumber \\
& = \zeta_k^T S_3 \zeta_k,
\end{align}
where $S_3$ is given in Appendix \ref{appd:matxSproc}.

Applying the S-procedure, if the LMI $G \triangleq A_a^T P A_a - E_a^T P E_a +U_1 +U_2 -S_1 - S_2 -S_3 < 0$
is feasible then $\Delta V(x_k) < 0$ is satisfied for the specific class of feedback-connected nonlinearities
$\phi \in \Phi_{sb}^{[0,\xi]} \cap \Phi_{sr}^{[0,\mu]}$.
All of introduced matrix (decision) variables are of compatible dimensions. $\square$
\medskip
\begin{rem}
The upper bounds on the integral terms in (\ref{eq:del_lyap1}) are derived by considering the a.e.\
continuity of the nonlinearity $\phi(\cdot)$. In those upper bounds the slope-restricted properties
of the nonlinear feedback are exploited, which gives sharper bounds than the upper bounds in \cite{HadB94b,HadK95,KapH96}.
\end{rem}

\subsection{Discrete-time Lur'e Systems with Slope-restricted and Odd Monotonic Nonlinearities}

Theorem \ref{stab:sufflyap_sr} exploits more information on the nonlinear operator than the original Lur'e problem.
This section derives a less conservative sufficient stability condition for the more restrictive class of nonlinearities that are odd monotonic,
by the introduction of additional quadratic constraints.
\begin{stab}
\label{stab:sufflyap_sbsr_odd}
Consider a system of the form (\ref{eq:1}) with the memoryless nonlinearity $\phi \in \Phi_{sb}^{[0,\xi]} \cap \Phi_{sr}^{[0,\mu]} \cap \Phi_{odd}$, which is continuous almost everywhere. A sufficient condition for global asymptotic stability is the existence
of a positive semidefinite matrix $P=P^T$ with a positive definite submatrix $P_{11}=P_{11}^T$ and diagonal positive semidefinite matrices
$Q$, $\tilde{Q}$, $T$, $\tilde{T}$, $N$, $L$, and $\tilde{L} \in \mathbb{R}^{n_q \times n_q}$ such that
\begin{equation}
\bar{G} =
\begin{bmatrix}
\bar{G}_{11} & \bar{G}_{12} & \bar{G}_{13} \\
\bar{G}_{12}^T  & \bar{G}_{22} & \bar{G}_{23} \\
\bar{G}_{13}^T & \bar{G}_{23}^T & \bar{G}_{33}
\end{bmatrix}
< 0
\label{eq:G2}
\end{equation}
where
\begin{align*}
\bar{G}_{11} & = G_{11}
\\
\bar{G}_{12} & = G_{12} - (CA - C)^T L - (CA + C)^T \tilde{L}
\\
\bar{G}_{13} & = G_{13} - (CA - C)^T L - (CA - C)^T \tilde{L}
\\
\bar{G}_{22} & = G_{22} - (CB - D)^T L - L (CB -D)
\nonumber \\
&
- (CB + D)^T \tilde{L} - \tilde{L} (CB + D)
\\
\bar{G}_{23} & = G_{23} - (CB - D)^T L - L D - \xi^{-1} L
\nonumber \\
&
- (CB -D)^T \tilde{L} - \tilde{L} D + \xi^{-1} \tilde{L}
\\
\bar{G}_{33} & = G_{33} - D^T L - L D - \tilde{L} D - D^T \tilde{L}
\end{align*}
\end{stab}
\noindent
{\em Proof:}
The difference in the Lyapunov function (\ref{eq:lyap_general}) between the $k+1$ and $k$ sampling instances is the same as $\Delta V(x_{k})$ in (\ref{eq:del_lyap1}) with the same definitions of $\zeta_k$, $A_a$, and $E_a$.
Sector-bounded, slope-restricted, odd-monotonic feedback-connected nonlinearities satisfy
(\ref{eq:sect_ineq}) and (\ref{eq:slope_ineq}) at each sampling time $k$ and all indices
$i=1,\ldots,n_q$ for $\phi \in \Phi_{sb}^{[0,\xi]} \cap \Phi_{sr}^{[0,\mu]} \cap \Phi_{odd}$ and satisfy
\begin{align}
(q_{k+1,i}-q_{k,i})[\phi_{k+1,i} + \phi_{k,i}] - \phi_{k,i}\frac{1}{\xi_i} \phi_{k+1,i} & \geq 0,
\label{eq:odd_ineq1}\\
(q_{k+1,i}-q_{k,i})[\phi_{k+1,i} + \phi_{k,i}] - \phi_{k,i}\frac{1}{\xi_i} \phi_{k+1,i} & \nonumber \\
\leq 2 q_{k+1,i} \phi_{k+1,i}. &
\label{eq:odd_ineq2}
\end{align}
The following notations are motivated by (\ref{eq:odd_ineq1}) and (\ref{eq:odd_ineq2}):
\begin{align}
& \sum_{i=1}^{n_q} 2 L_{ii} \left\{ \phi_{k,i} \xi_i^{-1}\phi_{k+1,i} -(q_{k+1,i}-q_{k,i})(\phi_{k+1,i} + \phi_{k,i}) \right\} \nonumber \\
& =
\zeta_k^T S_4 \zeta_k,
\end{align}
and
\begin{align}
& \sum_{i=1}^{n_q} 2 \tilde{L}_{ii}
\left\{
\begin{array}{c}
q_{k,i}(\phi_{k+1,i} - \phi_{k,i}) - q_{k+1,i}(\phi_{k+1,i} + \phi_{k,i})
\\
- \phi_{k,i} \frac{1}{\xi_i} \phi_{k+1,i}
\end{array}
\right\} \nonumber \\
&
=
\zeta_k^T S_5 \zeta_k,
\end{align}
where $S_4$ and $S_5$ are given in Appendix \ref{appd:matxSproc}.

If the LMI $G = A_a^T P A_a - E_a^T P E_a + U_1 + U_2 -S_1 - S_2 -S_3 - S_4 -S_5 < 0$ is feasible
then $\Delta V(x_k) < 0$ is satisfied for the class of feedback-connected nonlinearities
$\phi \in \Phi_{sb}^{[0,\xi]} \cap \Phi_{sr}^{[0,\mu]} \cap \Phi_{odd}$.
All of matrix (decision) variables introduced are of compatible dimensions. $\square$
\medskip
\begin{rem}
The inequality constraints over the odd monotonic nonlinearities (\ref{eq:odd_ineq1}) and (\ref{eq:odd_ineq2})
can be found in~\cite{NarT73,ThaS67,ThaSR67}. The inequalities (\ref{eq:odd_ineq1}) and (\ref{eq:odd_ineq2})
are written in terms of quadratic functions of $\zeta_k$, so that the S-procedure is applied to combine
the constraints with the negative definite condition over $\Delta V(x_k)$.
\end{rem}
%
%
\begin{rem}(Frequency-domain inequalities)
The well-known KYP lemma replaces an LMI with an equivalent frequency-domain inequality (FDI) in terms
of the system transfer function~\cite{Wil71}. The FDI provides a graphical tool to determine feasibility
of the LMI for systems with one nonlinearity, i.e., $n_p = n_q =1$, such as used in the graphical implementation of the Popov stability criterion.
It is straightforward to apply the KYP lemma to transform each LMI (\ref{eq:G1}) and (\ref{eq:G2}) into an FDI (not shown for space considerations).
\end{rem}
%

Theorem \ref{stab:sufflyap_sr} and \ref{stab:sufflyap_sbsr_odd} described LMI feasibility problems.
That is, each criterion considers whether the set $\{X \in \mathbb{S}^n|\Psi(X)<0,X \geq 0\}$ is empty or non-empty.
The following lemma shows the equivalence between two sets of matrix decision variables.
\begin{lem}(Feasibility considerations)
\label{lem:feasNonStrictStrict}
The set of symmetric matrices affine in $X$, $\{X \in \mathbb{S}^n|\Psi(X)<0,X \geq 0\}$, is nonempty if and only if
the set of symmetric matrices, $\{X \in \mathbb{S}^n|\Psi(X)<0,X > 0\}$ is nonempty.
\end{lem}
\noindent
{\em Proof:}
The (only if) part is obvious. To prove the (if) part, the LMIs in the set can be rewritten as:
\begin{equation}
\Psi(X) = F_0 + \sum_{i=1}^{N} x_i F_i ,
\end{equation}
where $X = X^T \in \mathbb{R}^{n \times n}$, $N = {n(n+1)}/{2}$, and the $F_i$ are of compatible dimension.
Suppose that $X^0 \in \{X \in \mathbb{S}^n|\Psi(X)<0,X \geq 0\}$. With the definition $X^{\delta} \triangleq X^0+\delta I$,
$X^{\delta}>0$ for any value of positive scalar $\delta >0$.
Now show that there exists a scalar $\delta >0$ such that $X^{\delta}$
is in the set $\{X \in \mathbb{S}^n|\Psi(X)<0,X > 0\}$. Consider
\begin{equation}
\Psi(X^{\delta}) =  F_0 + \sum_{i=1}^{N} (x_i+\delta_i) F_i ,
\end{equation}
where $\delta_i \in \{0,\delta \}$ for each $i=1,\ldots, N$ and the number of nonzero $\delta_i$ is $n$.
Since the eigenvalues of $\Psi(X^{\delta})$ are continuous for $\delta \rightarrow 0$,
these eigenvalues approach the eigenvalues of $\Psi(X^0)$ as $\delta \rightarrow 0$.  In particular, for sufficiently
small $\delta$ the eigenvalues of $\Psi(X^{\delta})$ have negative real part so that $\Psi(X^{\delta})<0$.
$\square$
\medskip
\begin{rem}(Computation for strict and non-strict LMIs)
From the results of Lemma \ref{lem:feasNonStrictStrict}, any LMI solver that is guaranteed
to converge for an LMI feasibility problem with strict inequalities will converge for the
above LMI feasibility problems.
\end{rem}
%
%
%
%
%
\section{Numerical Examples for Stability Analysis}
Numerical examples are provided to illustrate the analysis results in Section \ref{sec:suffstabmain}.
All LMI computations were performed with \textit{MATLAB's LMI Control Toolbox}.
The sector bounds $\xi_i$ and slope restrictions $\mu_i$ in each element $\phi_i$ of the nonlinearities were
taken to be the same.
The maximum upper bound on the $\xi$ was computed, where $\mu$ was taken to be linear in $\xi$, as shown in each example.
A large enough value for $\xi$ makes the LMI in Theorem \ref{stab:sufflyap_sr} (or \ref{stab:sufflyap_sbsr_odd}) infeasible.
A small enough value for $\xi$ makes the LMI feasible provided that the system is nominally stable, that is, the system without
nonlinearities is stable.
These two values provide upper and lower bounds for the value of $\xi$ for which the LMI switches from being feasible to being feasible.
The maximum $\xi$ for which the LMI in each criterion is feasible were computed by the bisection method.

As discussed in Section \ref{sec:suffstabmain}, the proposed stability criteria were derived for a Lyapunov function that has more degrees of freedom
than past stability criteria and so have the potential for being less conservative.
Table 1 shows that the criteria in Theorems 1 and 2 are less conservative for six numerical examples.
As expected, the stability margins are larger when more information is provided on the nonlinearities connected with the LTI system,
and Theorems \ref{stab:sufflyap_sr} and \ref{stab:sufflyap_sbsr_odd} provide more accurate estimates of the upper bound on $\xi$ or $\mu$.
Although Theorems \ref{stab:sufflyap_sr} and \ref{stab:sufflyap_sbsr_odd} apply to semiproper and strictly proper systems,
the numerical examples only include strictly proper systems to allow comparison to the other criteria in the literature,
which assume $D \equiv 0$.

\newcounter{MYtempeqncnt}

\begin{table*}[!t]
\setcounter{MYtempeqncnt}{\value{equation}}

\caption{The Maximal Upper Bound on the Sector Bound\label{tbl:ex1}}
\begin{center}
\begin{tabular}{|r |ccccccccccc|}
\hline
Stability Criterion    &
Ex $1$  & & Ex $2$  & & Ex $3$  & & Ex $4$  & &  Ex $5$ & & Ex $6$ \\
\hline
Circle $(\phi \in \Phi_{sb}^{[0,\xi]})$   & 1.0273    & & 0.18358      & & 0.21792   & &   2.91387    & & 0.03660  & & 0.03716  \\
Tsypkin $(\phi \in \Phi_{sb}^{[0,\xi]})$   & 1.0273    & & 0.18358       & & 0.21792      & &  2.91387       & & 0.03660 & & 0.03716  \\
Haddad et al.\ $(\phi \in \Phi_{sb}^{[0,\xi]} \cap \Phi_{sr}^{[0,\mu]})$  & 1.0273    & &0.18358& & 0.21792      & &  2.91387       & & 0.03660   & & 0.03716   \\
Kapila et al.\ $(\phi \in \Phi_{sb}^{[0,\xi]} \cap \Phi_{sr}^{[0,\infty]})$  & 1.0273  & & 0.18358& & 0.21792   & &  2.91387      & & 0.03660 & & 0.03716  \\
Park et al.\ $(\phi \in \Phi_{sb}^{[0,\xi]} \cap \Phi_{sr}^{[0,\infty]})$  & 1.7252    & & 0.18358 & & 0.21792   & &  2.91387      & & 0.03660 & & 0.03716  \\
Theorem 1 $(\phi \in \Phi_{sb}^{[0,\xi]} \cap \Phi_{sr}^{[0,\mu]})$   & 2.4475    & & 0.73082    & &  0.30203     & &  43.40412    & &  19.18289   & & 0.04613 \\
Theorem 2 $(\phi \in \Phi_{sb}^{[0,\xi]} \cap \Phi_{sr}^{[0,\mu]} \cap \Phi_{odd})$   & 2.5576    & & 0.73082       & & 0.83686   & & 43.40412    & & 19.18289 & & 0.18975 \\
\hline
\end{tabular}
\end{center}
\end{table*}

Example 1 is from \cite{ParK98} ($\mu = 2 \xi$):
\begin{equation*}
G(z) =
\frac{-0.5 z + 0.1 }
{(z^2 - z + 0.89)(z+0.1)}
\end{equation*}
Example 2 has 5 states and 2 nonlinearities ($\mu = \xi$):
\begin{align*}
& A=
\begin{bmatrix}
    0.2948  &       0 &        0  &       0 &        0 \\
         0  &  0.4568 &        0  &       0 &        0 \\
         0  &       0 &   0.0226  &       0 &        0 \\
         0  &       0 &        0  &  0.3801 &        0 \\
         0  &       0 &        0  &       0 &  -0.3270
\end{bmatrix}, \\
& B=
\begin{bmatrix}
   -1.1878  &  0.2341 \\
   -2.2023  &  0.0215 \\
    0.9863  & -1.0039 \\
   -0.5186  & -0.9471 \\
    0.3274  & -0.3744
\end{bmatrix}, \\
& C=
\begin{bmatrix}
   -1.1859 &   1.4725 &  -1.2173 &  -1.1283 &   -0.2611 \\
   -1.0559 &   0.0557 &  -0.0412 &  -1.3493 &   0.9535
\end{bmatrix}.
\end{align*}
Example 3 has a dense $A$-matrix and 2 nonlinearities ($\mu = \xi$):
\begin{align*}
& A=
\begin{bmatrix}
    0.0469  & -0.3992 &  -0.0835\\
    0.3902  & -0.5363 &  -0.2744\\
    0.4378  & -1.3576 &   0.4651
\end{bmatrix},\\
& B=
\begin{bmatrix}
   -0.5673 &  -0.2785\\
    0.1155 &  -0.0649\\
   -2.1849 &  -0.5976
\end{bmatrix}, \\
& C=
\begin{bmatrix}
    0.3587  & -1.0802 &  -0.6802\\
   -1.3833  & -1.0677 &   1.1497
\end{bmatrix}.
\end{align*}
Example 4 has two poles at the same location ($\mu = 2\xi$):
\begin{align*}
& A=
\begin{bmatrix}
    0.4030 &        0  &       0\\
         0 &  -0.1502  &       0\\
         0 &        0  & -0.1502
\end{bmatrix},
B=
\begin{bmatrix}
   -0.2494\\
    0.2542\\
   -0.2036
\end{bmatrix}, \\
& C=
\begin{bmatrix}
    0.9894 &   0.6649  &  0.4339
\end{bmatrix}.
\end{align*}
Example 5 has three poles at the same location ($\mu = 2\xi$):
\begin{align*}
& A=
\begin{bmatrix}
    0.4783 &        0 &        0 &        0\\
         0 &   0.7871 &        0 &        0\\
         0 &        0 &   0.7871 &        1\\
         0 &        0 &        0 &   0.7871
\end{bmatrix},\\
& B=
\begin{bmatrix}
   -1.5174\\
    1.2181\\
    0.2496\\
   -0.5181
\end{bmatrix}, \\
& C=
\begin{bmatrix}
   0.8457 &   -2.0885 &   1.2190  &  0.1683
\end{bmatrix}.
\end{align*}
Example 6 has a wide range of pole locations including two poles near 1 ($\mu = \xi$):
\begin{align*}
& A =
\mbox{diag}
\{
0.5359,0.9417,0.9802,0.5777,-0.1227,\\
& \qquad \quad \quad
-0.0034, -0.5721,0.2870,-0.3599
\},
\\
& B =
\begin{bmatrix}
     1  &   0  &   0  &   0\\
     0  &   1  &   0  &   0\\
     0  &   0  &   1  &   0\\
     0  &   0  &   0  &   1\\
     1  &   0  &   0  &   0\\
     0  &   1  &   0  &   0\\
     0  &   0  &   1  &   0\\
     0  &   0  &   0  &   1\\
     1  &   0  &   0  &   0
\end{bmatrix},
\\
& C =
\begin{bmatrix}
     1  &   1 &    0 &    0  &   0 &    0  &   0  &   0  &   0 \\
     0  &   0 &    1 &    1  &   1 &    0  &   0  &   0  &   0 \\
     0  &   0 &    0 &    0  &   0 &    1  &   1  &   0  &   0 \\
     0  &   0 &    0 &    0  &   0 &    0  &   0  &   1  &   1
\end{bmatrix}.
\end{align*}

The stability margins for Example 5 indicate that Theorem 1 can reduce the conservatism compared to literature results by more than 53,200\%.
The stability margins for Example 6 indicate that, for odd nonlinearities, Theorem 2 can reduce the conservatism by more than 400\% compared
to Theorem 1 and even more when compared to literature results.


\section{Conclusions}

This paper considers the analysis of global asymptotic stability for discrete-time Lur'e systems with nonlinearities of various types.
The proposed Lyapunov function (\ref{eq:lyap_general}) is a superset of all past reported Lyapunov functions, and the new sufficient conditions for globally asymptotic stability are less conservative.
Conservatism is further reduced for nonlinear operators that satisfy additional constraints in input and output behavior, such as being odd or having local slope restrictions. Numerical examples indicate that the reduction in conservatism can be many orders of magnitude. The results can be generalized to the analysis of $L_2$-gain, root-mean-square gain, and decay rate in the standard way \cite{BGFB94}.




\begin{figure*}[!t]
\normalsize


%

\appendices

\section{Definitions for Some Classes of Nonlinear Operators}
\label{appd:nonlinearftn}

\begin{align*}
\Phi_{sb}^{[0,\xi]}
& \triangleq
  \{\phi:\mathbb{R}^{n_q} \rightarrow \mathbb{R}^{n_q} \,| \,
  \phi_i(\sigma)\left[ \xi_i^{-1} \phi_i(\sigma) - \sigma \right] \leq 0, \, \forall \sigma \in \mathbb{R} \mbox{ and for } i=1,\ldots,n_q  \}; \\
\Phi_{sr}^{[0,\mu]}
& \triangleq
  \{\phi:\mathbb{R}^{n_q} \rightarrow \mathbb{R}^{n_q} \, | \,
  0 \leq \frac{\phi_i(\sigma)-\phi_i(\hat{\sigma})}{\sigma-\hat{\sigma}} \leq \mu_i, \, \forall \sigma \neq \hat{\sigma} \in \mathbb{R} \mbox{ and for } i=1,\ldots,n_q  \}; \\
\Phi_{odd}
& \triangleq
  \{\phi:\mathbb{R}^{n_q} \rightarrow \mathbb{R}^{n_p} \,|  \,
  \phi_i(\sigma) = -\phi_i(-\sigma), \, \forall \sigma \in \mathbb{R} \mbox{ and for } i=1,\ldots,n_q  \}.
\end{align*}
%

\section{Matrices for the Application of the S-Procedure in Theorems \ref{stab:sufflyap_sr} and \ref{stab:sufflyap_sbsr_odd}}
\label{appd:matxSproc}

\begin{align*}
U_1
& \triangleq
\begin{bmatrix}
0 & (CA-C)^T Q & -(CA-C)^T Q \\
\ast & Q(CB-D)+ (CB-D)^T Q - Q \mu^{-1} & QD + Q \mu^{-1} \\
\ast & \ast & -Q D -D^T Q - Q \mu^{-1}
\end{bmatrix} \\
U_2
& \triangleq
\begin{bmatrix}
A^T C^T \tilde{Q} \xi CA - C^T \tilde{Q} \xi C
& A^T C^T \tilde{Q} \xi C B + (CA-C)^T \tilde{Q} - C^T \tilde{Q} \xi D
& A^T C^T \tilde{Q} \xi D  \\
\ast
&
\begin{pmatrix}
B^T C^T \tilde{Q}\xi C B - D^T \tilde{Q} \xi D\\
 + (CB-D)^T \tilde{Q} + \tilde{Q} (CB-D) - \mu^{-1} \tilde{Q}
\end{pmatrix}
&  \mu^{-1} \tilde{Q} + \tilde{Q}D + B^T C^T \tilde{Q} \xi D \\
\ast & \ast & - \mu^{-1} \tilde{Q} + D^T \tilde{Q} \xi D
\end{bmatrix} \\
S_1
& \triangleq
\begin{bmatrix}
0 & C^T T & 0 \\
\ast & 2 \xi^{-1} T + T D + D^T T & 0 \\
\ast & \ast & 0
\end{bmatrix},
\quad
S_2 \triangleq
\begin{bmatrix}
0 & 0 & A^T C^T \tilde{T} \\
\ast & 0 & B^T C^T \tilde{T} \\
\ast & \ast & 2 \xi^{-1} \tilde{T} + \tilde{T}D + D^T \tilde{T}
\end{bmatrix} \\
S_3
& \triangleq
\begin{bmatrix}
0 & -(CA -C)^T  N &  (CA -C)^T  N  \\
\ast &  2N \mu^{-1} -(CB - D)^T N - N(CB - D) & -2N \mu^{-1} + (CB - D)^T N - N D\\
\ast & \ast & 2 N \mu^{-1} + D^T N + N D
\end{bmatrix} \\
S_4
& \triangleq
\begin{bmatrix}
0 & (CA - C)^T L &  (CA - C)^T L  \\
\ast &  (CB - D)^T L + L (CB - D) & (CB - D)^T L + L D + \xi^{-1}L \\
\ast & \ast & D^T L + L D
\end{bmatrix} \\
S_5
& \triangleq
\begin{bmatrix}
0 & (CA + C)^T \tilde{L} &  (CA - C)^T \tilde{L}  \\
\ast &  (CB + D)^T \tilde{L} + \tilde{L} (CB + D) & (CB - D)^T \tilde{L} + \tilde{L} D - \xi^{-1} \tilde{L}\\
\ast & \ast & D^T \tilde{L} + \tilde{L} D
\end{bmatrix}.
\end{align*}

\hrulefill
\vspace*{4pt}
\end{figure*}

%




%

\addtolength{\textheight}{-20cm}   

\end{document}